\documentclass[12pt]{iopart}

\begin{document}

\paper{Ideal Constraints - A Warning Note}

\author{Antonio S de Castro}

\address{UNESP - Campus de Guaratinguet\'a - Caixa Postal 205 - 12500000
Guaratinguet\'a - SP - Brasil}

\begin{abstract}
A criticism against the conception adopted by some textbook's authors
concerning ideal constraints is presented. The criticism is strengthen with
two traditional examples.

\vspace{15cm}

\noindent  Portuguese version  published in Revista Brasileira de Ensino
de F\'{\i}sica 22, 444 (2000): V\'{\i}nculos Ideais: Uma Nota de
Esclarecimento.

\end{abstract}

\maketitle

Constraints are restrictions that limit the motion of the particles of a
system. The forces necessary to constrain the motion are said to be forces
of constraint. The constraints whose unknown forces of constraint can be
eliminated are called ideal constraints. The principle of virtual work and
D'Alambert's principle are of paramount importance in pedagogical
presentations of analytical mechanics because they are readily derived from
Newton's laws and because they enable us to get rid of the unknown forces of
constraint from the equations of motion as well. Furthermore, these
principles are starting points for obtaining Lagrange's equations of motion.
To obtain the principle of virtual work and D'Alambert's principle
textbook's authors quite generally consider the virtual work done on a
system of $N$ particles: 
\[
\delta W=\sum_{i=1}^{N}\vec{F}_{i}^{(e)}\cdot \delta \vec{r}%
_{i}+\sum_{i=1}^{N}\vec{f}_{i}\cdot \delta \vec{r}_{i}
\]
where the force on each particle is written as the externally applied force $%
\vec{F}_{i}^{(e)}$ plus the force of constraint $\vec{f}_{i}$. The
principle of virtual work is applied only to static problems whereas
D'Alambert's principle is applied to dynamical situations. The difference
between these principles is not highlighted in this note neither is relevant
to the discussion which follows because the attention is focused on the
desembarrassment of the unwanted forces of constraint. Hauser \cite 
{h} argues that ``\textit{If the $\delta \vec{r}_{i}$'s are chosen so that
any constraints which exist between the coordinates of the particles are
satisfied, the constraint forces $\vec{f}_{i}$ acting on the particles will
be perpendicular to the displacements $\delta \vec{r}_{i}$}.'' Similarly
Lanczos \cite{l} argues that ``\textit{The vanishing of this scalar product
means that the force $\vec{f}_{i}$ is perpendicular to any possible virtual
displacement.}'' Taylor \cite{t} also argues that ``\textit{One thing known
about otherwise unknown forces of constraint and that is that they always
act at right angles to any conceivable displacement consistent with the
constraint under the condition of ``stopped time'', i.e., to any virtual
displacement.''} In the same line of reasoning Chow \cite{c} claims that ``%
\textit{Most of the constraints that commonly occur, such as sliding motion
on a frictionless surface and rolling contact without slipping, do no work
under a virtual displacement, that is} 
\[
\vec{f}_{i}\cdot \delta \vec{r}_{i}=0\;\;\;\;(4.8)
\]
\textit{This is practically the only possible situation we can imagine where
the forces of constraint must be perpendicular to $\delta \vec{r}_{i}$;
otherwise, the system could be spontaneously accelerated by the forces of
constraint alone, and we know that this does not occur...}''.

Indeed, the statements cited above hold for a system consisting of just one
particle. Nevertheless, there is no compelling reason to believe that they
are  true for a system with more than one particle, and they are not indeed.
Let us illustrate this point with two very instructive and traditional
problems: the rigid body and the Atwood machine.

A rigid body is a system of particles connected in such a way that the
distance between any two particles is invariable. Newton's third law implies
that for any pair of particles the forces of constraint are equals and
opposites and besides they are parallel to the relative position vector,
whatever the virtual displacements. These facts ensure that the net virtual
work of the forces of constraint vanishes.

In the Atwood machine two particles are connected by an inextensible string
passing over a pulley. If the string and the pulley are massless and the
motion is frictionless then the forces of constraint will reduce to the
tension in the string. The virtual displacements of the particles compatible
with the constraint will be in the vertical direction and so will the forces
of constraint. The virtual works of the forces of constraint on each
particle are the same, unless a sign, ensuring that the net virtual work
done by the forces of constraint vanishes.

It is worthwhile to observe that the conclusions obtained through the
former examples do not depend of the state of movement of the particles,
{\it i.e.}, if it is a static or a dynamical problem, in this way such
conclusions are suitable to the principle of virtual work as well as to
D'Alambert's principle.

From the previous two examples one can drawn the lesson that in order to
eliminate the forces of constraint is solely required that the net virtual
work vanishes: 
\[
\sum_{i=1}^{N}\vec{f}_{i}\cdot \delta \vec{r}_{i}=0
\]
This less restrictive condition allows forces of constraint not
perpendicular to $\delta \vec{r}_{i}$, for systems with more than one
particle,  without implying in spontaneous accelerated motion.

In short, there is absolutely no need for resorting to erroneous
restrictions on the forces of constraint, as those ones presented by Hauser,
Lanczos, Taylor and Chow, in order to eliminate them from the analytical
formulations of the classical mechanics. Ideal constraints are those which
the net virtual work on the entire system is zero whatever the relative
orientation among the forces of constraint and the virtual displacements.
This proviso is sufficient enough to ensure that the system is not
spontaneously accelerated. \newline

\end{document}